\documentclass[a4paper,12pt]{article}
\pdfoutput=1

\usepackage{times}
\usepackage[T1]{fontenc}
\usepackage[cp1250]{inputenc}
\usepackage{amsfonts}
\usepackage{amsmath}
\usepackage{graphics}
\usepackage{color}
\usepackage{epsfig,epsf}

\usepackage[colorlinks=true,linktocpage=true,linkcolor=blue,citecolor=blue]{hyperref}

\newcommand{\be}{\begin{equation}}
\newcommand{\ee}{\end{equation}}
\newcommand\beq{\begin{eqnarray}}
\newcommand\eeq{\end{eqnarray}}
\def\as{\alpha_{\mathrm{s}}}
\def\abar{\bar\alpha_{\mathrm{s}}}

\def\Nc{N_{\mathrm{c}}}

\begin{document}

\titlepage

\begin{center}
\vspace*{2cm}
{\Large \bf  	
Twist decomposition of exclusive heavy meson production cross sections
} \\
\vspace*{7mm}
Mariusz Sadzikowski
\\
\vspace*{7mm}
{\it
Jagiellonian University, Institute of Theoretical Physics,\\
\L{}ojasiewicza 11,  30-348 Krak\'{o}w, Poland}
\\[1em]
{\it 
}
\vspace*{5mm}

\today

\end{center}

\vspace*{2ex}

\begin{abstract}
We study the twist decomposition of the total cross sections for exclusive heavy vector meson electroproduction and photoproduction in the $\gamma^\ast p$ processes, within the leading logarithmic $1/x$ BFKL formalism. The Mellin transforms of the impact factors of the vector meson are calculated. We show that the higher twist contributions are strongly suppressed in the low-$x$ kinematical regime. Possible enhancement of the higher twists effects for nuclei targets is discussed.
\end{abstract}
\newpage
\section{Introduction}
\label{introduction}

Exclusive heavy vector meson electro- and photoproduction processes, already measured in many experiments \cite{H1:1996kyo,H1:1996gwv,ZEUS:1998cdr,H1:2000kis,ZEUS:2004yeh,H1:2005dtp,ZEUS:2009asc,H1:2013okq}, still planned for further future explorations \cite{FCC:2018byv,LHeC:2020van,AbdulKhalek:2021gbh}, are excellent probes of the gluon density at the low-$x$ corner of the phase space \cite{Jones:2015nna,Jones:2016ldq,Jones:2016icr}. In the low-$x$ regime the dipole picture provides a particularly effective description. In the nucleon (nuclei) rest frame, the virtual photon $\gamma^\ast$ fluctuates into the heavy quark-antiquark dipole, which interacts with the target and subsequently recombines into the heavy meson. These three stages are well separated in the dipole lifetime. An interesting question then arises of how the low-$x$, gluon dense regime contributes to the higher twist effects compared to the leading twist-2 description provided by the collinear factorization \cite{Collins:1996fb}. In this paper, we study the twist decomposition of the total cross sections for exclusive heavy vector meson production in the leading logarithmic $1/x$ LL BFKL formalism \cite{Kuraev:1977fs,Balitsky:1978ic}. We show that the higher twist contributions are strongly suppressed in the low-$x$ region, which results from the gluon reggeization process - the same as in the study of the DIS scattering in \cite{Motyka:2014jpa}.
It is worth noting here that this is in contrast to the strong enhancement prediction based on the multiple dipole scattering approximation \cite{Bartels:2000hv,Bartels:2009tu}. 

Chapter 2 contains main definitions, and new formulas for Mellin transform of vector meson impact factors are presented. Chapter 3 describes the BFKL dipole cross section evolution and main numerical results. In the last chapter, apart from the conclusions, one can find comments on the heavy vector meson production on nuclei and unitarity corrections to the BFKL evolution.

\section{Cross sections and impact factors}

The total cross sections for $\gamma^\ast p\rightarrow Vp$, where $V$ stands for $J/\Psi,\Upsilon$ particles, in the color dipole framework, can be described as \cite{Kowalski:2006hc}
\be
\label{tot_cross_section}
\sigma_{T,L}^{\gamma^\ast p\rightarrow Vp}(Q,y) = \int d^2rH_{T,L}(r,Q)\sigma_{q\bar{q}}(r,y),\;\; H_{T,L}(r,Q)=\sum_{f}\int\frac{dz}{4\pi}(\psi^\ast_V\psi)^f_{T,L}
\ee
where $T,L$ denotes the transverse and longitudinal polarizations of vector mesons of mass $M_V$. The $H_{T,L}$ are the corresponding impact factors
\beq
(\psi^\ast_V\psi)^f_{T} &=& \frac{Z_fe\Nc}{\pi z\bar{z}}\{m_f^2K_0(\epsilon_f r)\Phi_{T}(r,z)-[z^2+\bar{z}^2]\epsilon_f K_1(\epsilon_f r)\partial_r\Phi_{T}(r,z) \},\\
(\psi^\ast_V\psi)^f_{L} &=& \frac{Z_fe\Nc}{\pi}2Q( z\bar{z})K_0(\epsilon_f r)\left\{ M_V\Phi_{L}(r,z)+\delta\frac{m_f^2-\nabla^2_r}{M_Vz\bar{z}}\Phi_{L}(r,z) \right\}, 
\eeq
where $\nabla^2_r=\partial^2_r+1/r\partial_r$, $\epsilon_f^2 = Q^2z\bar{z} + m_f^2$ and $\bar{z}=1-z$. The $\Phi_{T,L}$ are wave functions that describe the heavy mesons. The sum of flavors in (\ref{tot_cross_section}) for all these processes is dominated by a single heavy quark ($f=c,b$), the charm of the mass $m_c$ and the fractional electric charge $Z_c=2/3$ in the case of $J/\Psi$, and the bottom of the mass $m_b$ and $Z_b=1/3$ in the case of $\Upsilon$. The dipole cross section $\sigma_{q\bar{q}}(r,y)$ is a function of the dipole size $r$, the rapidity $y=\ln(x_0/x)$ measured with respect to some initial value of the Bjorken variable $x_0$, and can be modeled using different approaches. In this letter, we focus on the BFKL evolution \cite{Kuraev:1977fs,Balitsky:1978ic}. In general, it is a function of the dimensionless variable $rQ_0$ where $Q_0$ is a saturation scale at the initial rapidity.

The twist decomposition is based on the Mellin transform technique \cite{Bartels:2000hv,Bartels:2009tu}, where the total cross section (\ref{tot_cross_section})  can be written in the Mellin space as
\be
\label{tot_cross_section_Mellin}
\sigma_{T,L}^{\gamma^\ast p\rightarrow Vp}(Q,y) = \int_C \frac{ds}{2\pi i} \tilde{H}_{T,L}(-s,Q)\tilde{\sigma}_{q\bar{q}}(s,y)
\ee 
where
\beq
\tilde{\sigma}_{q\bar{q}}(s,y) = \int_0^\infty d(r^2 Q_0^2) (r^2 Q_0^2)^{s-1} \sigma_{q\bar{q}}(rQ_0 ,y),\;
\tilde{H}_{T,L}(-s,Q) = \int d^2r (r^2 Q_0^2)^{-s} H_{T,L}(r,Q) .
\eeq
For vector meson wave-functions we used two models \cite{Kowalski:2006hc}:
$$
\Phi^{(i)}_{T,L} = N^{(i)}_{T,L}z\bar{z}\varphi^{(i)}_{T,L}(z)\exp (-r^2h^{(i)}(z)/2R^{(i)\,2}_{T,L}),\;\; i=1,2
$$
where $\varphi^{(1)}_T(z) = z\bar{z}, \varphi^{(1)}_L(z) = 1, h^{(1)}(z)=1, R^{(1)}_{T,L}\equiv R_{T,L}$, $\delta=0$ for the so called Gauss-LC model and $\varphi^{(2)}_T(z) =\varphi^{(2)}_L(z) = \exp\{-m_f^2{\cal R}^2/(8z\bar{z})+m_f^2{\cal R}^2/2\}, h^{(2)}(z)=4z\bar{z}, R^{(2)}_T=R^{(2)}_L\equiv {\cal R}$, $\delta=1$ for the so called boosted-Gauss model. Then the Mellin transform of the impact factors takes the form
\beq
\label{imp_fact_Q}
\tilde{H}^{(i)}_{L}(-s,Q) &=& \frac{Ze\Nc N^{(i)}_L}{\pi}\int_0^1 dz\varphi^{(i)}(z)z\bar{z}\frac{QM_V}{\epsilon_f^2} \left(\frac{4Q_0^2}{\epsilon_f^2}\right)^{-s}\left(u^{(i)}_L\right)^{1-s}
\left\{\left[z\bar{z}+\frac{\delta m_f^2}{M_V^2}\left(1+\frac{2h^{(i)}}{m^2R_L^2}\right)\right]\right. \nonumber \\
&\cdot& \left.\Gamma(1-s)^2 U(1-s,1,u^{(i)}_L) - \frac{2\delta h^{(i)}}{M_V^2 R^{(i)\,2}_T}\Gamma(2-s)^2 U(2-s,1,u^{(i)}_L) \right\},  \nonumber \\
\tilde{H}^{(i)}_{T}(-s,Q) &=& \frac{Ze\Nc N^{(i)}_T}{2\pi}\int_0^1 dz\varphi^{(i)}(z)\frac{m_f^2}{\epsilon_f^2} \left(\frac{4Q_0^2}{\epsilon_f^2}\right)^{-s}\left(u^{(i)}_T\right)^{1-s}
\left\{\Gamma(1-s)^2 U(1-s,1,u^{(i)}_T) \right. \nonumber \\
&+& \left. \frac{2(z^2+\bar{z}^2)}{m_f^2 R^{(i)\,2}_T}h^{(i)}(z)\Gamma(1-s)\Gamma(2-s) U(1-s,0,u^{(i)}_T) \right\},\;\; u^{(i)}_{T,L}=\frac{\epsilon_f^2 R^{(i)\,2}_{T,L}}{2h^{(i)}(z)}
\eeq
where $\Re s < 1$. For the vector meson photoproduction only the transverse cross section survives and in the case of the Gauss-LC model the compact form of the impact factor is available:
\beq
\label{impact_factor_Q0}
\tilde{H}^{(1)}_{T}(-s,Q^2=0) &=& \frac{Ze\Nc N^{(1)}_T}{12\pi}\left(\frac{4Q_0^2}{m_f^2}\right)^{-s}\Gamma(1-s)
(\frac{m_f^2 R_T^2}{2})^{1-s}\left\{ \Gamma(1-s)U(1-s,1,\frac{m_f^2 R_T^2}{2}) \right. \nonumber \\
&+& \left. \frac{6}{5m_f^2 R_T^2}\Gamma(2-s)U(1-s,0,\frac{m_f^2R_T^2}{2}) \right\}
\equiv \left(\frac{4Q_0^2}{m_f^2}\right)^{-s} f^{(1)}_T(s,m_f R_T)
\eeq
where $U(1-s,a,u)$ are the Tricomi confluent hypergeometric functions.
There is no simple analytical form for the boosted-Gauss model; however, the general structure of the impact factor $\tilde{H}^{(2)}_{T}(-s,Q^2=0) = (4Q_0^2/m_f^2)^{-s}f^{(2)}_T(s,m_f {\cal R})$ remains the same. It should be noted that the functions $f^{(i)}_T$ are analytical for $\Re s<1$ and can be expanded in the inverse powers of $m_fR$ based on the relation:
\be
\label{series}
u^{1-s}U(1-s,a,u) = \sum_{k=0}^\infty \frac{C_k^{(a)}(s)}{u^k}
\ee
where the lowest coefficients are $C_0^{(0)}=1,  C_1^{(0)} = -(1-s)(2-s), C_0^{(1)} = 1, C_1^{1)} = -(1-s)^2$. This expansion can be very useful as the analytical approximation, in particular for bottom quarks, where the first three terms give the exact result within 1 per cent. 

\section{Twist expansion in the BFKL framework}

The leading BFKL rapidity evolution of the dipole cross section in the Mellin space is given by \cite{Lipatov:1996ts}
\be
\label{sigma_qq}
\tilde{\sigma}_{q\bar{q}}(s,y) = -\sigma_0 \Gamma(s)\exp\{\abar \chi(s) y\},\;\;\chi(s)=2\psi(1)-\psi(-s)-\psi(1+s)
\ee
where $\abar=N_c \as/\pi$ is a strong coupling constant, $\psi$ is digamma function, and the initial condition $\sigma_{q\bar{q}}(r,y=0) = \sigma_0(1-\exp\{-r^2Q_0^2\})$
suggested by the Golec-Biernat - W{\"u}sthoff model was adopted \cite{Golec-Biernat:1998zce,Golec-Biernat:1999qor}. With this choice, the Mellin
strip in (\ref{tot_cross_section_Mellin}) is located inside the range $-1<\Re s<0$. Note that the dipole scattering
formula brings the simple poles from the initial condition and essential singularities from the BFKL evolution at negative integer
values of the parameter $s$. At the same time, as mentioned above, the impact factors $f^{(i)}$ are analytic functions for $\Re s<0$. Therefore, one can exchange the Mellin strip with the sum over the poles located at negative integers
\be
\label{sigma_twist}
\sigma_{T}^{\gamma^\ast p\rightarrow Vp}(0,y) = \sum_{n=1}^\infty
\int_{C_{-n}} \frac{ds}{2\pi i}  \left(\frac{4Q_0^2}{m_f^2}\right)^{-s} f^{(i)}_T(s,m_f R^{(i)}_T)\tilde{\sigma}_{q\bar{q}}(s,y),
\;\; C_{-n}: |s+n|=\epsilon
\ee

\begin{figure}[b]
 \begin{center}
 \includegraphics[width=7.5cm]{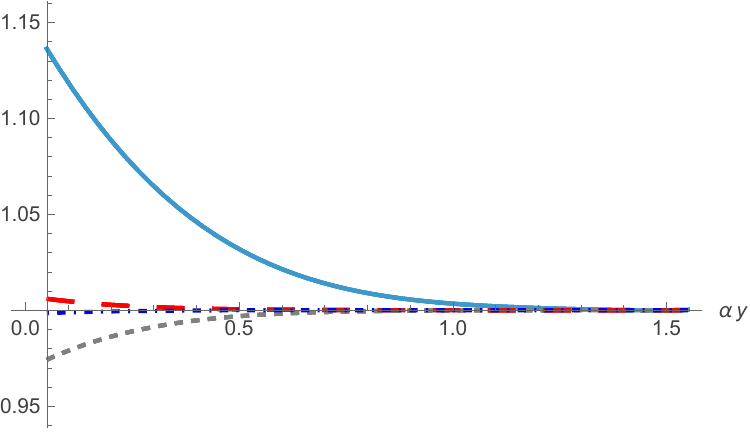}\includegraphics[width=7.5cm]{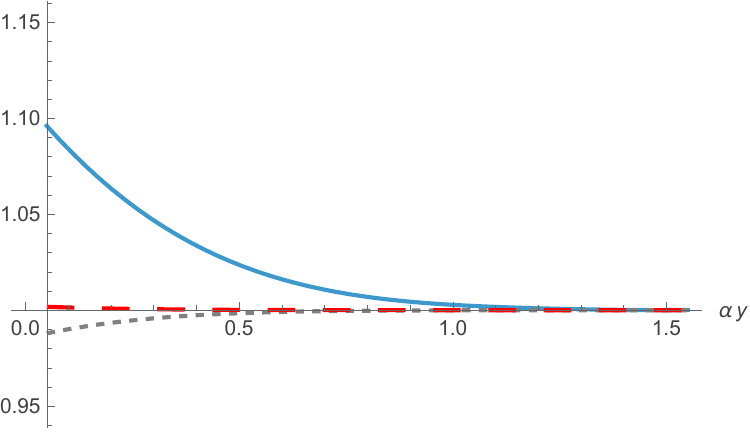}
 \caption{Relative aggregate contributions of the twists to the transverse total cross section of $J/\Psi$ photoproduction as a function of $\abar y$: $K_1$ (solid line), $K_2$ (dashed line), $K_3$ (long dashed line), $K_4$ (dotted-dashed line). Left panel describes the Gauss-LC model the right panel boosted-Gaussian model.}
   \label{sigmar}
 \end{center}
\end{figure}
\noindent
for any small $\epsilon$. The above equality must be understood as an asymptotic series expansion \cite{Bartels:2000hv,Bartels:2009tu}.
This is the basic relation for the twist expansion in powers of $(4Q_0^2/m_f^2)^n$, where twist of the order $2n$ is given by the integral over a circle $C_{-n}$. It was already shown in \cite{Motyka:2014jpa} that for the process of deep-inelastic electron-proton scattering, higher twist contributions are suppressed with increasing rapidity. The same effect is also present in the case of exclusive heavy meson electroproduction and photoproduction. For better understanding of the effect, it is convenient to explicitly separate the pole structure in the BFKL eigenvalue $\chi(-n+\gamma) = 1/\gamma-2H_{n-1}+\chi^{(n)}_{reg}(\gamma)$ for any natural number $n$ and a small value of $|\gamma|<\epsilon$, where $H_n$ is a harmonic function ($H_0\equiv 0$), and $\chi^{(n)}_{reg}(\gamma)$ is a regular function vanishing at $\gamma=0$. Using this decomposition and equation (\ref{sigma_twist}) one can write
\beq
\label{sigma_twist2}
&&\sigma_{T}^{\gamma^\ast p\rightarrow Vp}(0,y) = \sum_{n=1}^\infty \left(\frac{4Q_0^2}{m_f^2}\right)^{n}
e^{-2H_{n-1}\abar y} A_n(y) \equiv \sum_{n=1}^\infty \sigma_{T,2n}^{\gamma^\ast p\rightarrow Vp}(0,y),\\
&&A_n(y)=-\sigma_0\int_{C_0} \frac{d\gamma}{2\pi i} \exp\left\{\abar y[1/\gamma+\chi^{(n)}_{reg}(\gamma)]+\gamma t\right\} \Gamma(-n+\gamma)f^{(i)}_T(-n+\gamma,m_f R^{(i)}_T) \nonumber
\eeq
where $t=\ln(m_f^2/4Q_0^2)$. The first line of equation (\ref{sigma_twist2}) explicitly shows the exponential suppression
of the higher twist terms with rapidity by the harmonic function, whereas the leading twist ($n=1$) remains intact. 
The dependence on the rapidity through $A_n$ is polynomial. It is important to remember that this result is in contrast to the eikonal approximation, where in the twist expansion the saturation scale $Q_0$ increases with increasing rapidity.

\begin{figure}[t]
 \begin{center}
 \includegraphics[width=7.5cm]{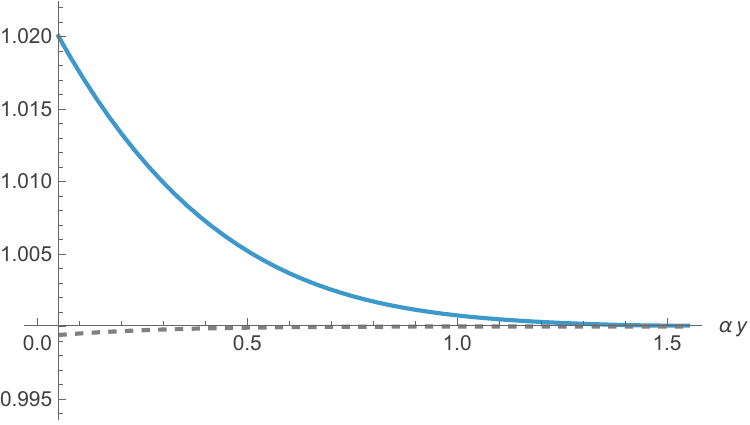}\includegraphics[width=7.5cm]{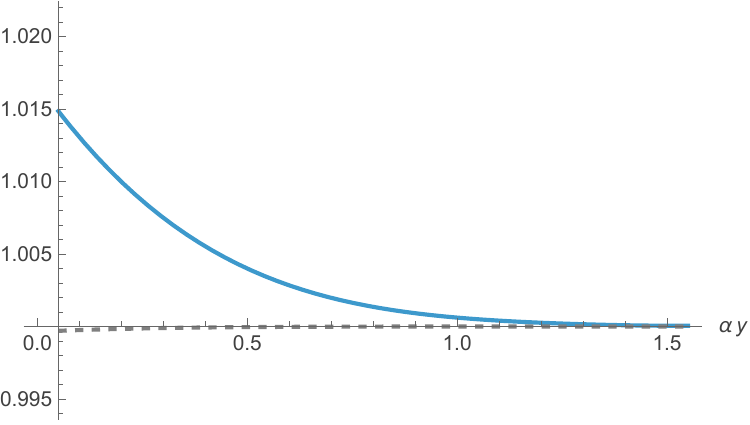}
 \caption{Relative aggregate contributions of the twists to the transverse total cross section of $\Upsilon$ photoproduction as a function of $\abar y$: $K_1$ (solid line), $K_2$ (dashed line). Left panel describes the Gauss-LC model the right panel boosted-Gaussian model. The first two twists reproduce the full cross section within 1 per mile.}
   \label{sigmar}
 \end{center}
\end{figure}

For presentation, we adopt the parameters values taken from \cite{Kowalski:2006hc}:
$m_c=1.4$ GeV, $R_T^2=6.5$ GeV$^{-2}$, $R_L^2=3$ GeV$^{-2}$, ${\cal R}^2=3.5$ GeV$^{-2}$ for $J/\Psi$ and \cite{Goncalves:2014swa}: $m_b=4.2$ GeV, $R_T^2=1.91$ GeV$^{-2}$, ${\cal R}^2=0.57$ GeV$^{-2}$ for $\Upsilon$. Normalizations $N^{(i)}_{T,L}$ are irrelevant because we present only the relative aggregate twist contributions to the total cross section $K_n=\sum_{k=1}^n\sigma_{T,2k}^{\gamma^\ast p\rightarrow Vp}/\sigma_{T}^{\gamma^\ast p\rightarrow Vp}$. For dipole parameterization, it is enough to assume the initial saturation scale $Q_0=0.25$ GeV, which is appropriate for a nucleon target, and all results are shown as functions of $\abar y$. Therefore, we do not need to explicitly specify the values of the strong coupling constant and the initial value of $x_0$. Note that all the figures are presented in the range $\abar y\in (0.05,1.55)$ which, taken into account the typical values of the strong coupling at the heavy quark masses, corresponds numerically to the rapidity up to 5-7.
In Fig. 1, Fig. 2 we present relative aggregate contributions $K_n$ for $J/\Psi$ and $\Upsilon$ photoproductions. In the case of $J/\Psi$ particle, for rapidities larger than unity, the first twist describes the full cross section within 2-3 percent accuracy (depending on the model) which becomes exponentially better with increasing rapidity. For the particle $\Upsilon$, this precision is already below one per cent. This clearly predicts that the cross sections for the heavy meson photoproductions are mainly a leading twist effect at large rapidity (small $x$) region.

\begin{figure}[b]
 \begin{center}
 \includegraphics[width=7.5cm]{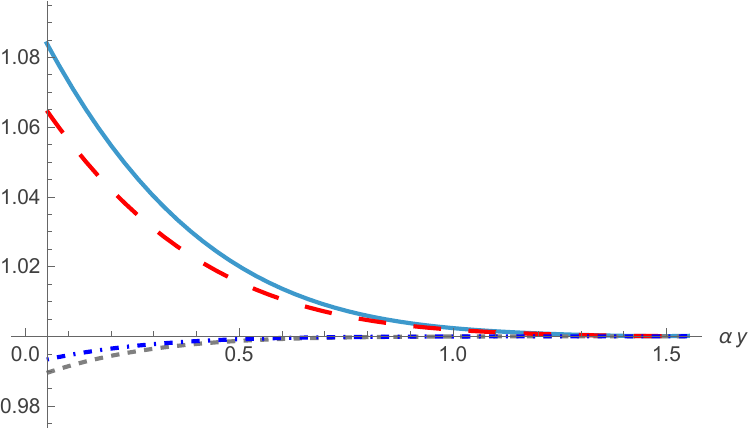}\includegraphics[width=7.5cm]{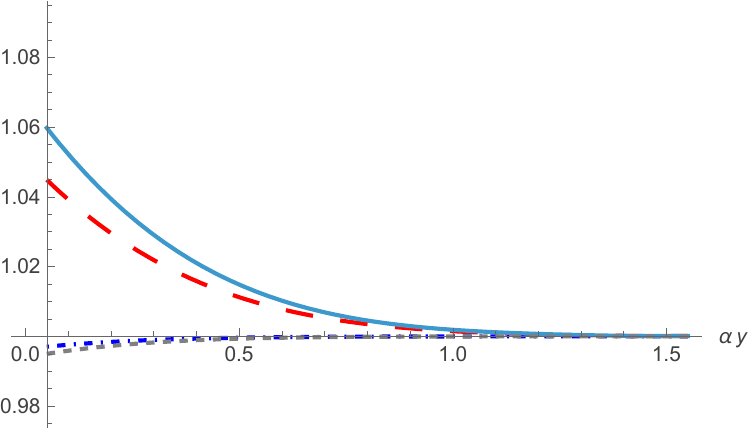}
 \caption{Relative aggregate contributions of the twists to the transverse total cross section of $J/\Psi$ photoproduction (left panel) and longitudinal total cross section (right panel) as a function of $\abar y$: $K_1$ (solid line), $K_2$ (dashed line) for photon virtuality $Q^2=3.5$ GeV$^{2}$ and $K_1$ (long dashed line), $K_2$ (dotted-dashed line) for photon virtuality $Q^2=5$ GeV$^{2}$. Results are obtained within the Gauss-LC model.}
   \label{sigmar}
 \end{center}
\end{figure}


At non-zero photon virtuality, the typical twist expansion parameter contains terms $\epsilon_f^2$ in the denominator (see equation (\ref{imp_fact_Q})), which, as expected, further suppress the higher twist contributions with growing $Q^2$. There is no simple analytical expression for the final result; however, different possible approximations are available, notably based on the expansion (\ref{series}). Even then, the final analytical formulae are not compact, but it shows that the twist expansion parameters are weighted with a combination of $4Q^2_0/m_f^2$ and $4Q_0^2/(Q^2+4m_f^2)$ scales. In the photoproduction limit, only the first scale survives. The general twist expansion is given by analog of equation (\ref{sigma_twist}) adapted for non-zero $Q^2$
\be
\label{tot_cross_section_expansion}
\sigma_{T,L}^{\gamma^\ast p\rightarrow Vp}(Q,y) = \sum_{n=1}^\infty\int_{C_{-n}} \frac{ds}{2\pi i} \tilde{H}_{T,L}(-s,Q)\tilde{\sigma}_{q\bar{q}}(s,y)\equiv \sum_{n=1}^\infty \sigma_{T,2n}^{\gamma^\ast p\rightarrow Vp}(Q,y)
\ee 
where $\tilde{H}_{T,L}$ and $\tilde{\sigma}_{q\bar{q}}$ are given by equations (\ref{imp_fact_Q}) and (\ref{sigma_qq}). The definition of relative aggregate contributions $K_n$ remains as before. In Fig. 3  the twist decomposition of the total transverse and longitudinal cross sections for $J/\Psi$ electroproduction at two photon virtualities $Q^2=3.5, 5$ GeV$^{2}$ is presented. Only the Gauss-LC model was used because the boosted-Gaussian model is qualitatively similar with around 10$\%$ stronger suppression. As is expected growing photon virtuality makes the leading twist approximation more accurate. In the case of $\Upsilon$ electroproduction, higher twists are safely negligible, as already were at zero photon virtuality.

\section{Summary and conclusions}

In this paper, we performed the twist decomposition of the cross sections for photo- and electroproduction of heavy mesons $J/\Psi$ and $\Upsilon$ on the nucleon target. Our main prediction is the suppresion of higher twists effects with growing rapidity. In the case of $J/\Psi$ the leading twist contribution describes the full cross section with accuracy of per cent value for rapidities of order $y\sim 2$ (Fig. 1). In the case of $\Upsilon$ the contributions with the higher twists are within the per mile region (Fig. 2). Heavy meson electroproduction with non-zero $Q^2$ makes the scale harder with stronger higher twist suppression compared to photoproduction processes (Fig. 3). 

\begin{figure}[t]
 \begin{center}
 \includegraphics[width=7.5cm]{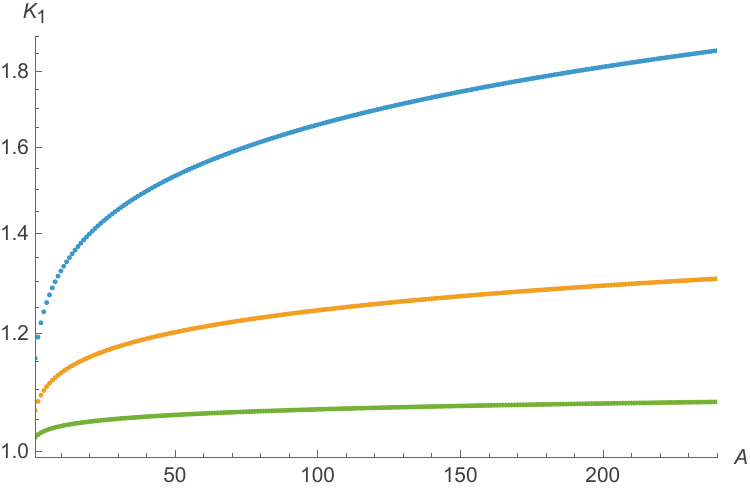}\includegraphics[width=7.5cm]{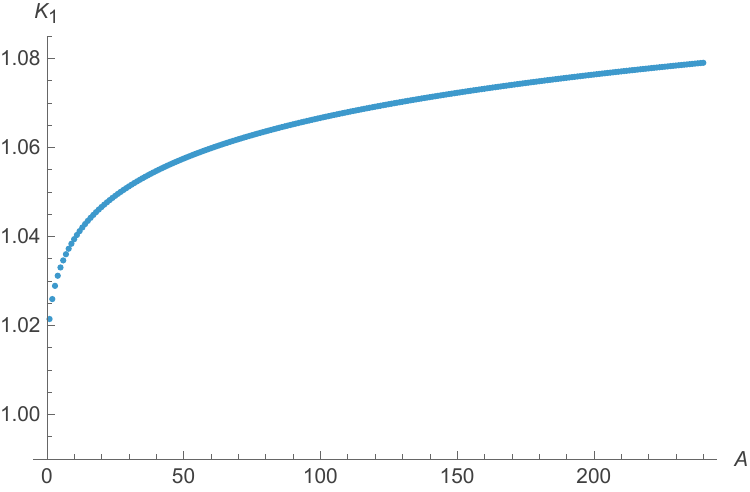}
 \caption{Leading twist contribution $K_1$ to the transverse total cross section of $J/\Psi$ photoproduction as a function of the atomic mass $A$ of the nuclei target. In the left hand panel a vertical scale is logarithmic and the rapidities are set to $\abar y = 0, 0.3, 0.6$ (from the top to the bottom). In the right hand panel the same variable in the linear scale for the rapidity $\abar y=0.6$. Results are obtained within the Gauss-LC model.}
   \label{sigmar}
 \end{center}
\end{figure}

It is very interesting to discuss the photoproduction of heavy mesons on nuclei with atomic number $A$. In this situation, the change in the saturation scale of the process is expected, according to the simple rule $Q_A^2\sim A^{1/3}Q^2_0$. Using this relation as an equality, Fig. 4 shows the parameter $K_1$ as a function of the atomic mass $A$ for $J/\Psi$ photoproduction, which provides us with an estimate of higher twist contributions. The atomic number span was checked to 240 which corresponds to the saturation scale $2Q_A/m_c\in (0.18,0.9)$ that is still inside the twist expansion domain. As is clearly seen from Fig. 4, the higher twists contributions can be substantially enhanced, however, the effect decreases with increasing rapidity (see the left hand panel of Fig. 4) because the effective saturation scale $Q_A$ is unable to overcome the higher twist suppression at small enough values of the variable $x$. Nevertheless, in the mid-rapidity region $\abar y=0.6$, the enhancement is reasonably large, starting from below 1 per cent in the case of nucleons to around 8 per cent for very heavy nuclei (see the right panel of Fig. 4). What is also interesting is that the curve is steep at low values of atomic masses and already for light nuclei ($A\sim 10-20$) the higher twist contributions triple in comparison to the nucleon case. For higher values of the rapidity, the curve is essentially flat.

It is well known that the BFKL evolution requires unitarity corrections, which is provided by the Balitsky-Kovchegov (BK) equation, valid in the large $N_c$ limit \cite{Balitsky:1995ub,Kovchegov:1999yj,Kovchegov:1999ua}. However, it was shown in the case of DIS scattering, that the non-linear corrections affect mainly the leading twist-2 contribution \cite{Motyka:2023pmt} with minor corrections to the higher twists. We believe that similar effect occurs in the case studied here, therefore, the main conclusions of the paper remain robust, although detailed studies are left for future work.



\section*{Acknowledgements}
The author thanks Leszek Motyka for several useful discussions.

\bibliographystyle{unsrt}
\bibliography{vmp}

\end{document}